# A First-principles approach to predict Seebeck coefficients: Application to $La_{3-x}Te_4$

Yi Wang[1], Yong-Jie Hu[1], Shun-Li Shang[1], Samad A. Firdosy[2], Kurt E. Star[2], Jean-Pierre Fleurial[2], Vilupanur A. Ravi[2,3], Long-Qing Chen[1], and Zi-Kui Liu[1]

[1]Department of Materials Science and Engineering, The Pennsylvania State University, University Park, PA 16802, USA

[2]Jet Propulsion Laboratory, California Institute of Technology. 4800 Oak Grove Drive, Pasadena, CA 91109, USA

[3]Department of Chemical and Materials Engineering, California State Polytechnic University, Pomona, CA 91768

Theoretical descriptions of the Seebeck coefficient in terms of the differential electrical conductivity given by Cutler and Mott is the foundation of later works in terms of transmission function from the thermoelectric transport theory. On the other hand, recent studies in the literature have shown the relation between the Seebeck coefficient and chemical potential of electrons. In this work, this relation is rigorously derived from fundamental thermodynamics, and an formalism for the parameter-free calculation of the Seebeck coefficient based on the electronic density-of-states from first-principles calculations is presented. Numerical results are given using the *n*-type $La_{3-x}Te_4$ thermoelectric material as the prototype. With the rigid band approximation, the calculated temperature dependences of the Seebeck coefficients of $La_{3-x}Te_4$ as a function of carrier concentration show excellent agreements with experimental data.

Thermoelectric materials [1-2] can be used to generate electricity, measure temperature or change the temperature of objects due to the thermoelectric effects which refer to the reversible (see the review by Wood [3]) Seebeck effect, Peltier effect, or Thomson effect, each of which deals with the direct conversion of temperature differences at dissimilar metal junctions to electric voltage and vice versa [4-7]. In particular the Seebeck coefficient represents the magnitude of an induced thermoelectric voltage in response to a temperature difference across a material. The Seebeck coefficient is also known as thermopower, thermoelectric power, or thermoelectric sensitivity and is defined as $\alpha = \Delta\phi/\Delta T$, representing the magnitude of an induced thermoelectric voltage, $\Delta\phi$, in response to a temperature difference, $\Delta T$, across a material. A modern thermoelectric device is composed of *p*-type semiconductor and *n*-type semi-conductors which are coupled to the heat source through a hot shoe and the heat sink through the cold shoe. While the theory of the thermoelectric effect appears to be well established and widely applied in the literature, the microscopic theory of thermoelectrics and the parameter-free calculation of the thermoelectric property remain challenging. This is particularly true for the calculation of the Seebeck coefficient. Earlier theoretical descriptions of the Seebeck coefficient in terms of the differential electrical conductivity was given by Cutler and Mott [8] which was the foundation of later works [9-15] in terms of the transmission function [16-19] from the thermoelectric transport theory [20-22]. Recent studies [23-25] have noticed the possible relation between the Seebeck coefficient and system's chemical potential or the electrochemical potential.

Hereby we show a procedure for the efficient calculation of the Seebeck coefficient based on the electronic density-of-states (e-DOS) calculated by the first-principles method without invoking any adjustable parameters. Our starting point is from the Mermin [26-27] statistics, from which the conservation equation for the total number of electrons is

$$\text{Eq. 1} \quad \int n(\varepsilon) f d\varepsilon = N$$

where $n(\varepsilon)$ is the electronic density-of-state (e-DOS), $\varepsilon$ the band energy, $N$ the total number of electrons in the system, and $f$ the Fermi-Dirac distribution

$$\text{Eq. 2} \quad f = \frac{1}{\exp\left[\dfrac{\varepsilon - \mu(T)}{k_B T}\right] + 1}$$

where $k_B$ is the Boltzmann's constant, $T$ the temperature, and $\mu(T)$ the chemical potential of electron. We note that the temperature dependence of $\mu(T)$ plays the central role for the thermoelectric effects as seen below.

A natural intuition is that the Seebeck effect is due to a thermoelectric electromotive force. Because there are no any moving parts in the system except the electrons, the change in chemical potential of electron must be the only reason behind the thermoelectric electromotive force. For a uniform material, only temperature change can result in the change of the chemical potential of electron. As a result, a thermoelectric electromotive force is related to the chemical potential of thermal electron, i.e., the temperature dependent portion of the free energy gain per electron.

$$\text{Eq. 3} \quad \phi = \mu(T) - \varepsilon_F(V)$$

where $\varepsilon_F(V)$ is the Fermi energy which is volume ($V$) dependent.

Accounting for both temperature and volume effects, the Seebeck coefficient should be the total derivative of $\phi$ with respect to $T$. Under constant pressure ($P$), we get

$$\text{Eq. 4} \quad \alpha_P = \frac{d\phi}{dT}$$

where the subscript $P$ indicates constant pressure.

Typically for the *n*-type semiconductor, the chemical potential of electrons at 0 K (Fermi energy) is located slightly above the bottom of the conduction band, as shown in Figure **1** for $La_3Te_4$ and $La_{2.75}Te_4$ in the plots of e-DOSs. For the insulator, the Fermi energy locates exactly at the top of the valence band, as shown in Figure **1** for $La_{2.67}Te_4$ in the plots of e-DOS. It is mentioned in the literature [35-36] that the temperature dependence of $\mu$ is the reason behind the Seebeck effect. Using $La_{2.75}Te_4$ as the demonstration case, the calculated temperature dependences of $\mu$ at the carrier concentration of $1.2\times10^{20}$ *e*/cm$^3$ (by shifting the Fermi energy, equivalent to removing 0.223 *electron* per formula of $La_{2.75}Te_4$) is plotted in Figure 2. The temperature dependence of $\mu$ is solely dictated by the behavior of the e-DOS by the present formalism, so is the Seebeck coefficient. The faster the change of the e-DOS in the vicinity of the Fermi energy with respect to the band energy, the faster of the change of $\mu$ with respect to temperature, and the larger the Seebeck coefficient. This is in agreement with, but not limited to, the concept of convergence band [2, 37]. Numerically, the rapid increase of the e-DOS with increasing band energy is the reason why $\mu$ decreases with increasing temperature, resulting in a negative Seebeck coefficient for $La_{3-x}Te_4$, due to the combined effects of Eq. 1 and Eq. 2.

The criteria for a good *n*-type semiconductor can be described as follows:

i) at 0 K, relatively low values of e-DOS at the Fermi energy which in turn is located at slightly above the bottom of the conduction band, as shown in the plot of the e-DOS for $La_{2.75}Te_4$ in Figure **1**; and

ii) the e-DOS increases rapidly with the increasing values for the electron band energy.

As a result, the Seebeck coefficient can be calculated directly with one-dimensional numerical integration. Knowing the fact that the e-DOS is a basic output of most modern first-principles codes, the present formulation makes it a lot easier to search for superior thermoelectric materials by means of high-throughput first-principles calculation [29-30].

Next, we detail the first-principles calculations of temperature dependences of the Seebeck coefficients for Lanthanum telluride ($La_{3-x}Te_4$) to demonstrate the proposed formalism. $La_{3-x}Te_4$ is used for thermoelectric power generation under the high temperature environment. A thermoelectric material is often characterized by the carrier concentration, i.e., the number of electrons in the conduction band (or the number of holes in the valence band) which are mostly implemented by doping the perfect crystal. In principle, a precise first-principles calculation should be performed using the doped structure. However, doing so is often very time consuming. An alternative solution is to adopt the rigid band approximation [22]. In this approach, the electronic band structure is first calculated for a referenced crystal structure. After that the electronic band structure is assumed to remains unchanged with only the Fermi energy is adjusted to fit the desired carrier concentrations. In order to study the effects of different referenced crystal structures on the calculated Seebeck coefficients, in the present work, we have considered three referenced crystal structures, with the compositions of $La_3Te_4$, $La_{2.75}Te_4$, and $La_{2.67}Te_4$.

From viewpoint of chemical valence, the cation La has a valence +2, and anion Te has a valence of -3. It can therefore be anticipated that vacancy at the La site can make the material transform from a metal at $x = 0$ into an insulator at $x = 1/3$, knowing the fact that $La_3Te_4$ has one electron located at the conduction band, and $La_{2.67}Te_4$ has no electron located at the conduction band. We consider a variety of carrier concentrations of $4.0 \times 10^{21}$, $3.8 \times 10^{21}$, $2.9 \times 10^{21}$, $2.0 \times 10^{21}$, $1.6 \times 10^{21}$, $4.4 \times 10^{20}$, $1.3 \times 10^{20}$, and $1.2 \times 10^{20}$ $e/cm^3$. These carrier concentrations correspond to the reduced Hall carrier concentrations of $\eta_H$ = 0.91, 0.87, 0.65, 0.45, 0.36, 0.10, 0.029, and 0.027, respectively, given in the measurements made by May et al. [38]. Effectively, $La_3Te_4$ corresponds to $\eta_H = 1$ and $La_{2.67}Te_4$ corresponds to $\eta_H = 0$.

We employed the projector-augmented wave (PAW) method [31-32] implemented in the Vienna *ab initio* simulation package (VASP, version 5.3) together with the Perdew-Burke-Ernzerhof revised for solids (PBEsol) [33] exchange-correlational functional. An energy cutoff of 219.3 eV has been used for the crystal structure relaxations and the calculations of the interatomic force constants while an energy cutoff of 284.7 eV has been used for the calculation of 0 K static energy. The thermal expansions have been calculated following the standard quasiharmonic phonon approach [34] with the calculation details and results to be reported in a separate work.

The calculated e-DOSs for the three structures are plotted in Figure **1**. Based on the calculated e-DOS, the different carrier concentrations can be implemented by positive doping (p-doping, removing electrons) of $La_3Te_4$, negative doping (n-doping, adding electrons) of $La_{2.67}Te_4$, or n-doping of $La_{2.75}Te_4$ for high carrier concentrations ($\eta_H$ = 0.91, 0.87, 0.65, 0.45, and 0.36) and p-doping of $La_{2.75}Te_4$ for low carrier concentrations $\eta_H$ = 0.10, 0.029, and 0.027.

The three sets of Seebeck coefficients calculated based on three referenced crystal structures of $La_3Te_4$, $La_{2.75}Te_4$, and $La_{2.67}Te_4$, are compared with the experimental data for $La_{3-x}Te_4$ from May et al. [38] superimposed in Figure **3**, Figure **4**, and Figure **5**, respectively. The modest deviations between the calculations and experiments for lower carrier concentrations at $\eta_H$ = 0.10, 0.029, and 0.027 can be in part attributed to the experimental difficulties, due to reasons such as sample inhomogeneity and oxidation etc [34]. This is particularly true as it is seen that the calculated difference at $\eta_H$ = 0.029 and 0.027 is one magnitude smaller than the measured one by May et al. [38]. The difference between the Seebeck coefficients at $\eta_H$ = 0.029 and 0.027 should not have been as large as that reported from the overall good agreements between the calculations and experiments in the whole carrier concentration range between 0.91 and 0.027. As discussed by May et al. [38], when approaching to the insulating limit of the stoichiometric $La_{2.67}Te_4$, the uncertainty associated with electrical resistivity and Seebeck coefficient is considerably large. It was seen that the measured Hall carrier concentrations showed ~10% uncertainties against the nominal vacancy concentration (i.e. the value of x in $La_{3-x}Te_4$) and the Hall carrier concentrations were slightly underestimated for larger x (i.e. small $\eta_H$).

In summary, we have presented a first-principles approach for the parameter-free prediction of the Seebeck coefficient solely based on the electronic density-of-states under the rigid band approximation. The effect of lattice thermal expansion is accounted for by quasiharmonic phonon approximation. Numerical results are given using the *n*-type high temperature thermoelectric material $La_{3-x}Te_4$ at *x*=0, 0.25, and 0.33 as the prototype. The predicted temperature dependences of the Seebeck coefficients of $La_{3-x}Te_4$ at the carrier concentration of $4.0 \times 10^{21}$, $3.8 \times 10^{21}$, $2.9 \times 10^{21}$, $2.0 \times 10^{21}$, $1.6 \times 10^{21}$, $4.4 \times 10^{20}$, $1.3 \times 10^{20}$, and $1.2 \times 10^{20}$ *e*/cm$^3$ show excellent agreements with experimental data.

**Acknowledgements**

This research was carried at the Jet Propulsion Laboratory, California Institute of Technology under a contract with the National Aeronautics and Space Administration and at the Pennsylvania State University under a subcontract. First-principles calculations were carried out partially on the LION clusters at the Pennsylvania State University supported by the Materials Simulation Center and the Research Computing and Cyberinfrastructure unit at the Pennsylvania State University, partially on the resources of NERSC supported by the Office of Science of the US Department of Energy under contract No. DE-AC02-05CH11231, and partially on the resources of XSEDE supported by NSF with Grant No. ACI-1053575.

**Figure captions**

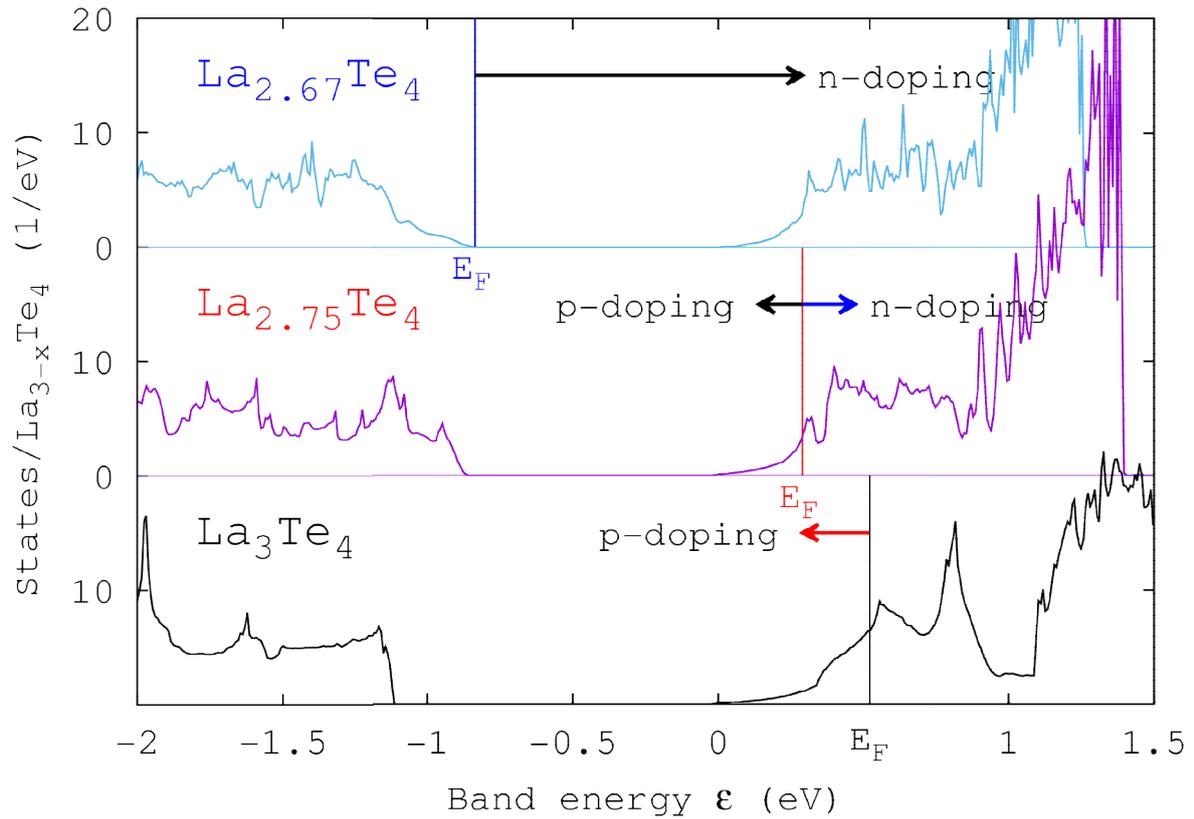

Figure 1. Calculated electronic density-of-states for La$_3$Te$_4$, La$_{2.75}$Te$_4$, and La$_{2.67}$Te$_4$. The vertical lines with label "E$_F$" indicates the Fermi energies without doping. The arrows label the possible types of doping for the three referenced compositions.

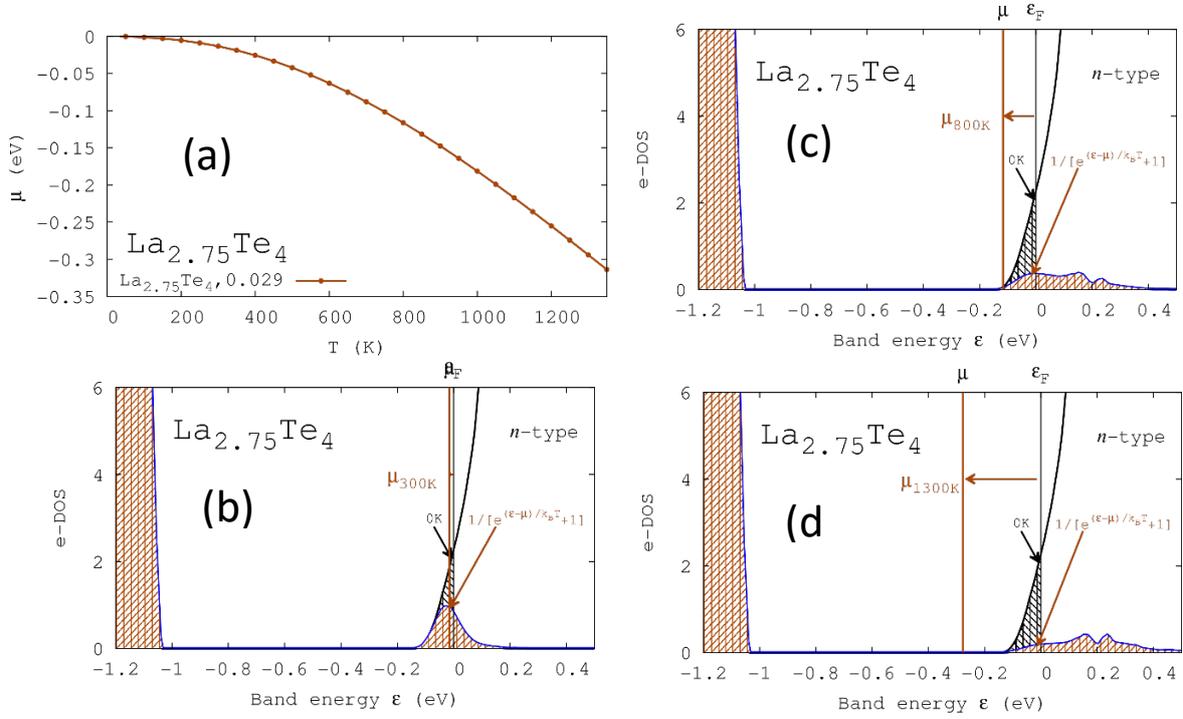

Figure 2. Calculated chemical potential (μ's) of electron for La$_{2.75}$Te$_4$ at the doping level that results in the carrier concentration of $1.2\times10^{20}$ $e$/cm$^3$, which correspond to the reduced Hall carrier concentrations of 0.027 given by the experiments of May et al. [38]. (a) as a function of temperature; (b) at 300 K focused at the Fermi energy with the upper filed pattern area representing the 0 K electronic occupation while the lower filled pattern area representing finite temperature electronic occupation by the Fermi distribution; (c) same as (b) except at 800 K; (d) same as (b) except at 1300 K.

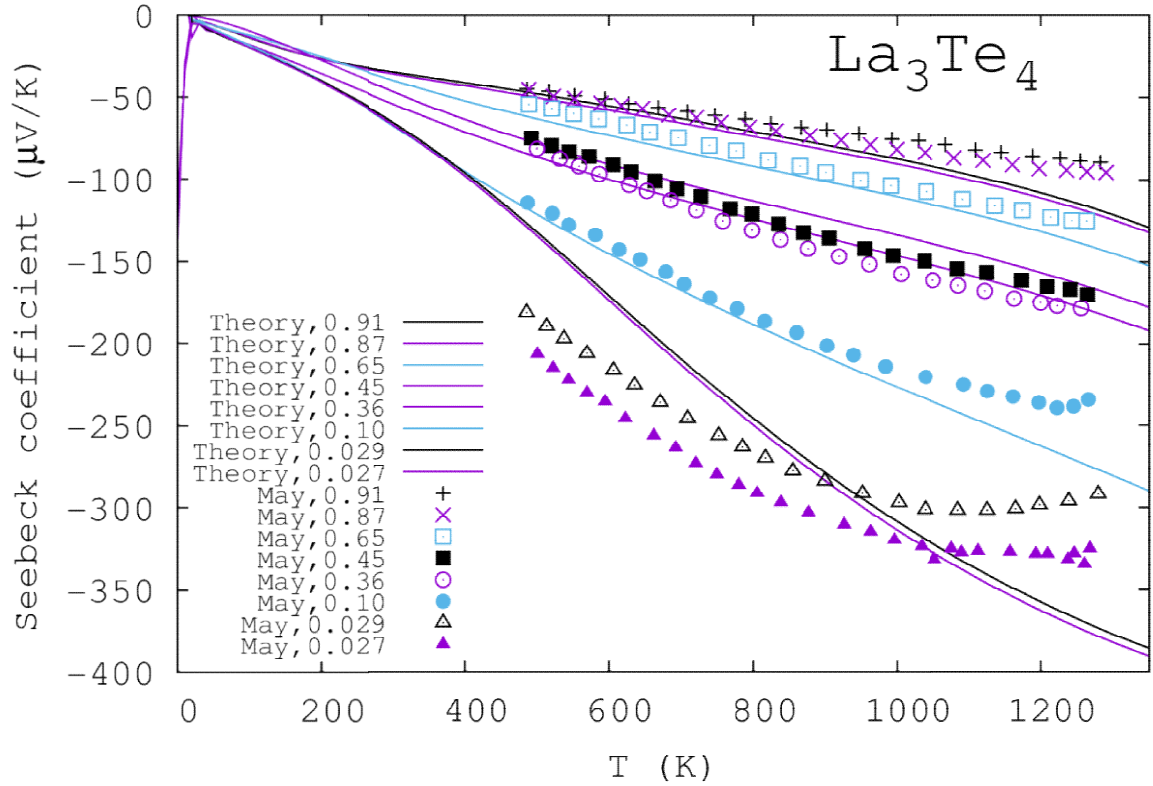

Figure 3. Calculated Seebeck coefficients for $La_{3-x}Te_4$ based on the electronic density-of-states of $La_3Te_4$. The carrier concentrations of $4.0\times10^{21}$, $3.8\times10^{21}$, $2.9\times10^{21}$, $2.0\times10^{21}$, $1.6\times10^{21}$, $4.4\times10^{20}$, $1.3\times10^{20}$, and $1.2\times10^{20}$ $e$/cm$^3$ correspond to the reduced Hall carrier concentrations of 0.91, 0.87, 0.65, 0.45, 0.36, 0.10, 0.029, and 0.027, respectively, given by the experiments of May et al. [38].

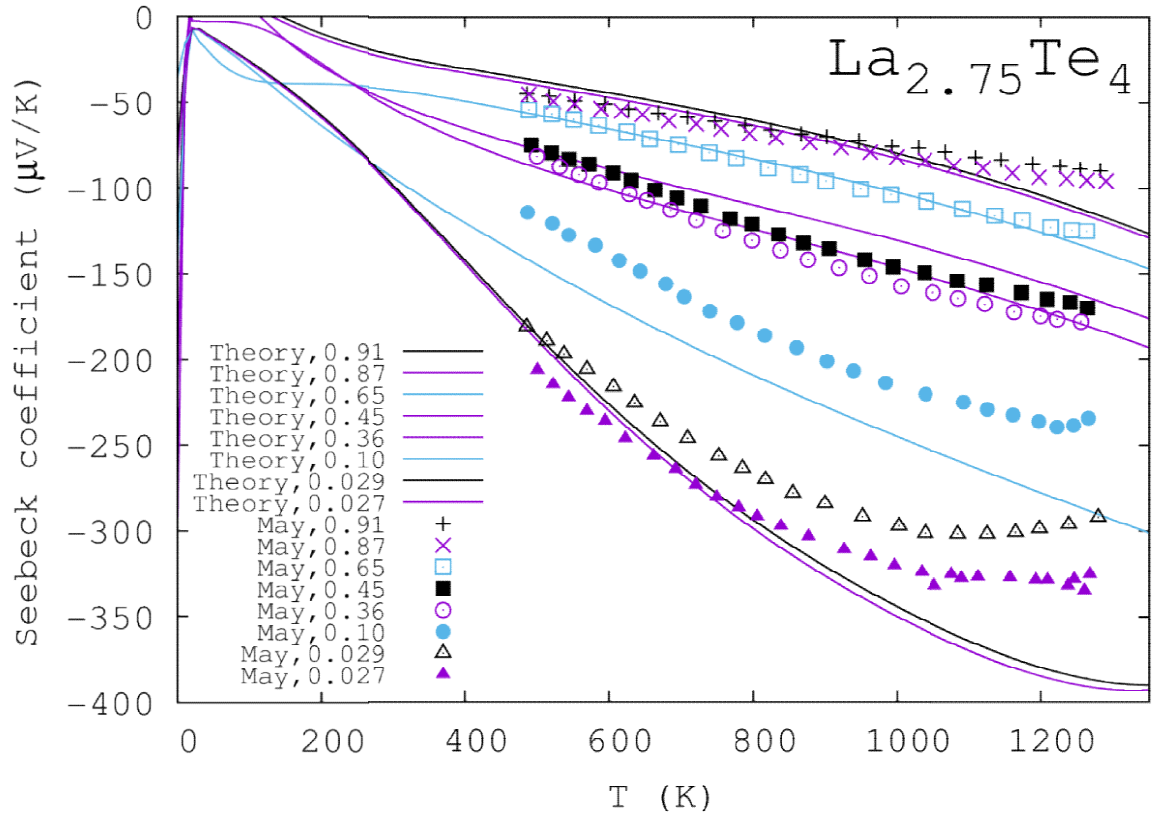

Figure 4. Calculated Seebeck coefficients for $La_{3-x}Te_4$ based on the electronic density-of-states of $La_{2.75}Te_4$. The carrier concentration of $4.0 \times 10^{21}$, $3.8 \times 10^{21}$, $2.9 \times 10^{21}$, $2.0 \times 10^{21}$, $1.6 \times 10^{21}$, $4.4 \times 10^{20}$, $1.3 \times 10^{20}$, and $1.2 \times 10^{20}$ $e$/cm$^3$ correspond to the reduced Hall carrier concentrations of 0.91, 0.87, 0.65, 0.45, 0.36, 0.10, 0.029, and 0.027, respectively, given by the experiments of May et al. [38].

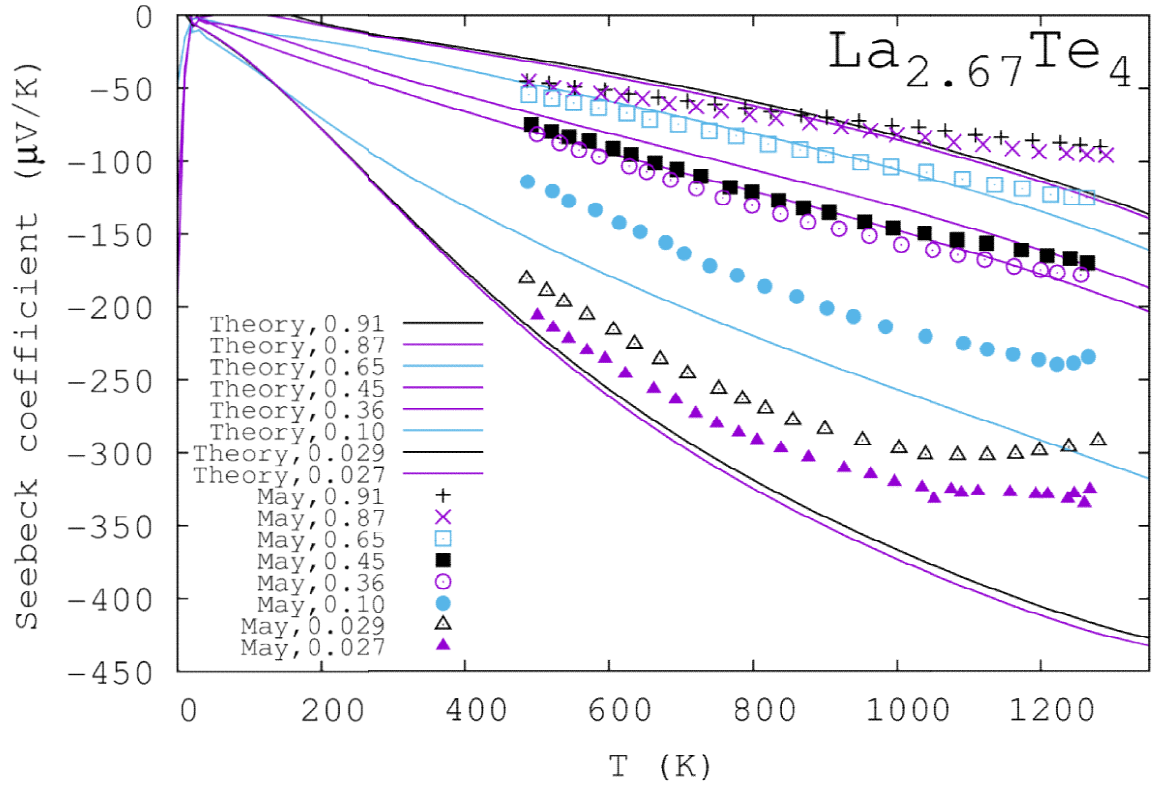

Figure 5. Calculated Seebeck coefficients for $La_{3-x}Te_4$ based on the electronic density-of-states of $La_{2.67}Te_4$. The carrier concentration of $4.0\times10^{21}$, $3.8\times10^{21}$, $2.9\times10^{21}$, $2.0\times10^{21}$, $1.6\times10^{21}$, $4.4\times10^{20}$, $1.3\times10^{20}$, and $1.2\times10^{20}$ $e$/cm$^3$ correspond to the reduced Hall carrier concentrations of 0.91, 0.87, 0.65, 0.45, 0.36, 0.10, 0.029, and 0.027, respectively, given by the experiments of May et al. [38].